\documentclass[cernpreprint, atlasdraft=false, UKenglish, texmf, orcidlogo]{atlasdoc}
 
\usepackage{atlaspackage}
\usepackage{siunitx}
\usepackage{placeins}
\usepackage{booktabs}
\usepackage{multirow}

\usepackage{atlasbiblatex}

\usepackage{atlasphysics}
\usepackage{atlasjetetmiss}
\usepackage{atlasparticle}
\usepackage{atlasmetadefs}
\usepackage{atlasmisc}
\usepackage{amsmath}
\usepackage{cleveref}

\ifthenelse{\boolean{AtlasDraft}}{\usepackage[output=true, shift=true]{}}{}
 
\graphicspath{{logos/}{figures/}}
 
\usepackage{ANA-JETM-2020-02-PAPER-defs}

\AtlasTitle{Performance and calibration of quark/gluon-jet taggers using 140~fb$^{-1}$ of $pp$ collisions at $\sqrt{s} = 13$~TeV with the ATLAS detector}

\AtlasAbstract{
The identification of jets originating from quarks and gluons, often referred to as quark/gluon tagging, plays an important role in various analyses performed at the Large Hadron Collider, as Standard Model measurements and searches for new particles decaying to quarks often rely on suppressing a large gluon-induced background. This paper
describes the measurement of the efficiencies of quark/gluon taggers developed within the ATLAS Collaboration, using $\sqrt{s} = 13$~TeV proton--proton collision data with an integrated luminosity of 140~fb$^{-1}$ collected by the ATLAS experiment. Two taggers with high performances in rejecting jets from gluon over jets from quarks are studied: one tagger is based on requirements on the number of inner-detector tracks associated with the jet, and the other combines several jet substructure observables using a boosted decision tree. A method is established to determine the quark/gluon fraction in data, by using quark/gluon-enriched subsamples defined by the jet pseudorapidity. Differences in tagging efficiency between data and simulation are provided for jets with transverse momentum between 500~GeV and 2~TeV and for multiple tagger working points.
}

\author{The ATLAS Collaboration}
 
\AtlasRefCode{JETM-2020-02}
 
\PreprintIdNumber{CERN-EP-2023-151}

\AtlasJournal{Chin. Phys. C}
\AtlasJournalRef{Chin. Phys. C 48 (2024) 023001}
\AtlasDOI{10.1088/1674-1137/acf701}

\hypersetup{pdftitle={ATLAS document},pdfauthor={The ATLAS Collaboration}}
 
\usepackage{chngcntr}
 
\begin{document}
 
\maketitle
 
\tableofcontents

\section{Introduction}
\label{sec:intro}
Various Standard Model (SM) measurements~\cite{ref1,ref8} and searches for physics beyond the SM~\cite{ref9} at the Large Hadron Collider (LHC)~\cite{Evans:2008zzb} benefit from the identification of showers of hadronic particles (jets) originating from quarks or gluons. Several searches employing quark/gluon ($q/g$) tagging, techniques that tag jets originating from a quark or a gluon, have demonstrated improved sensitivity to new physics and the ability to discriminate between new resonances that decay into different types of hadronic jets~\cite{ref-dijet,ref-dijet2,PhysRevD.91.055019,doi:10.1142/9789813231467_0032}. For example, in some supersymmetry scenarios, many final-state light quarks can be produced~\cite{ATLAS:2012qgw,ATLAS:2012dp}. The ability to discriminate between quark- and gluon-initiated jets, hereafter referred to as \enquote{quark-jets} and \enquote{gluon-jets},  therefore provides a powerful tool to use in searches for new physics. If a new particle were discovered, such a discriminant could provide valuable information about the nature of the particle. Moreover, accurate identification of the origin of jets is crucial in certain SM measurements, such as when reconstructing hadronic decays of $W$ bosons in a  measurement of the mass of the top quark.

In the theory of quantum chromodynamics (QCD), quarks and gluons are not free particles in their kinematic evolution, and they produce streams of particles that the LHC experiments can measure. Discrimination between jets of different partonic origins has been attempted at several experiments~\cite{CMS:2021iwu,ref22,CMS:2017yer,CMS:2013kfv,CMS:2017pcy,PhysRevD.98.092014,CMS:2018vzn,Altheimer:2012mn,Andersen:2016qtm,Larkoski:2014pca,CDF:2005prv,OPAL:1999jkz,CLEO:2007tqf,DELPHI:1999gah,ZEUS:2004gcp}. For example, some analyses~\cite{ref24,Komiske:2016rsd} used the full radiation pattern inside a jet as an image processed in a deep neural network classifier. Most work has relied on jet properties that result from the different colour charges of the partons. According to QCD, the colour charge of a gluon is larger than that of a quark by a factor of $9/4$~(the \enquote{Casimir ratio})~\cite{ALTARELLI1977298}. Hence, in their kinematic evolution and hadronisation, the gluons produce more particles, leading to jets with a higher number of constituents and a broader radiation pattern than in quark-jets. Advances in the theoretical~\cite{ref16} and phenomenological~\cite{ref17, ref18, Gras:2017jty, Komiske:2018vkc} understanding of the radiation patterns of quark- and gluon-jets have led to recent progress in $q/g$ tagging. Compared to previous studies which considered single-variable taggers for a lower \pt range~\cite{Aad_2014,ATL-PHYS-PUB-2017-009}, this study focuses on the construction of a new $q/g$ tagger that utilises several jet substructure variables, and on extending the $q/g$ tagging of jets to a higher energy range.

This paper investigates the performance of two $q/g$ taggers, which are built with the goal of identifying quark-jets and rejecting gluon-jets. The first tagger is based on a requirement placed on the charged-particle multiplicity (\ntrk) of a jet. The second tagger employs a boosted decision tree (BDT) that takes as input a set of jet kinematic and substructure variables. The $q/g$ tagging efficiencies are estimated in data by using a method that splits the data sample into subsamples where the fraction of quark-jets is higher or lower than the fraction of gluon-jets (i.e.\ quark- or gluon-enriched subsamples, respectively).
 
The paper is structured as follows. Section~\ref{sec:atlas} introduces the ATLAS detector. A brief description of the data and Monte Carlo (MC) samples used in the analysis is given in Section~\ref{sec:Data-MC}. In Section~\ref{sec:Obj-event} the object definitions and event selection criteria used to select events and classify them into the various categories are described. Variables used in the definition of the $q/g$ taggers studied in this analysis are presented in Section~\ref{sec:QG-var}. The method developed to evaluate the $q/g$ tagging efficiencies and the taggers' discrimination power is presented in Section~\ref{sec:QG-method}. Systematic uncertainties affecting the analysis are detailed in Section~\ref{sec:QG-syst}. Measurements of the tagging efficiencies in data and their ratio to those expected from MC simulation (scale factors) are shown in Section~\ref{sec:QG-SF}, while conclusions are drawn in Section~\ref{sec:conc}.

\section{ATLAS detector}
\label{sec:atlas}
 
\newcommand{\AtlasCoordFootnote}{
ATLAS uses a right-handed coordinate system with its origin at the nominal interaction point (IP)
in the centre of the detector and the \(z\)-axis along the beam pipe.
The \(x\)-axis points from the IP to the centre of the LHC ring,
and the \(y\)-axis points upwards.
Cylindrical coordinates \((r,\phi)\) are used in the transverse plane,
\(\phi\) being the azimuthal angle around the \(z\)-axis.
The pseudorapidity is defined in terms of the polar angle \(\theta\) as \(\eta = -\ln \tan(\theta/2)\).
Angular distance is measured in units of \(\Delta R \equiv \sqrt{(\Delta\eta)^{2} + (\Delta\phi)^{2}}\).
}
 
The ATLAS detector~\cite{PERF-2007-01} at the LHC covers nearly the entire solid angle around the collision point.\footnote{\AtlasCoordFootnote}
It consists of an inner tracking detector surrounded by a thin superconducting solenoid, electromagnetic and hadron calorimeters,
and a muon spectrometer incorporating three large superconducting air-core toroidal magnets.
 
The inner-detector system (ID) is immersed in a \qty{2}{\tesla} axial magnetic field
and provides charged-particle tracking in the range \(|\eta| < 2.5\).
The high-granularity silicon pixel detector covers the vertex region and typically provides four measurements per track,
the first hit normally being in the insertable B-layer, which was installed before Run~2~\cite{PIX-2018-001}.
It is followed by the silicon microstrip tracker, which usually provides eight measurements per track.
These silicon detectors are complemented by the transition radiation tracker (TRT),
which enables radially extended track reconstruction up to \(|\eta| = 2.0\).
The TRT also provides electron identification information
based on the fraction of hits (typically 30 in total) above a higher energy-deposit threshold corresponding to transition radiation.
 
The calorimeter system covers the range \(|\eta| < 4.9\). Within the region \(|\eta| < 3.2\), electromagnetic calorimetry is provided by barrel and
endcap high-granularity lead/liquid-argon (LAr) calorimeters,
with an additional thin LAr presampler covering \(|\eta| < 1.8\)
to correct for energy loss in material upstream of the calorimeters.
Hadron calorimetry is provided by the steel/scintillator-tile calorimeter,
segmented into three barrel structures within \(|\eta| < 1.7\), and two copper/LAr hadron endcap calorimeters.
The solid angle coverage is completed with forward copper/LAr and tungsten/LAr calorimeter modules
optimised for electromagnetic and hadronic energy measurements, respectively.
 
The muon spectrometer comprises separate trigger and
high-precision tracking chambers measuring the deflection of muons in a magnetic field generated by the superconducting air-core toroidal magnets.
The field integral of the toroids ranges between \num{2.0} and \qty{6.0}{\tesla\metre}
across most of the detector.
Three layers of precision chambers, each consisting of layers of monitored drift tubes, cover the region \(|\eta| < 2.7\). They are complemented by cathode-strip chambers in the forward region, where the background is highest.
The muon trigger system covers the range \(|\eta| < 2.4\) with resistive-plate chambers in the barrel and thin-gap chambers in the endcap regions.
 
Interesting events are selected by the first-level trigger system implemented in custom hardware,
followed by selections made by algorithms implemented in software in the high-level trigger~\cite{TRIG-2016-01}.
The first-level trigger accepts events from the \qty{40}{\MHz} bunch crossings at a rate below \qty{100}{\kHz},
which the high-level trigger reduces in order to record events to disk at about \qty{1}{\kHz}.
 
An extensive software suite~\cite{ATL-SOFT-PUB-2021-001} is used in data simulation, the reconstruction
and analysis of real and simulated data, in detector operations, and in the trigger and data acquisition
systems of the experiment.

\section{Data and simulation samples}
\label{sec:Data-MC}
 
The data used in this analysis were collected by the ATLAS detector between 2015 and 2018 from proton--proton collisions at a centre-of-mass energy of $\sqrt{s} = 13$~\TeV, and correspond to an integrated luminosity of 140~\ifb~\cite{ATLAS:2022hro1}. The events included in this dataset satisfy quality requirements which ensure that all detector systems were operational~\cite{ATLAS:2019fst}. This was achieved by monitoring detector-level quantities and the characteristics of reconstructed collision events at key stages of the data processing chain.
 
MC simulations are used to model SM multijet production, which is the main process expected to play a role in this analysis. The \PYTHIA[8.230]~\cite{Sjostrand:2014zea} generator was used to simulate multijet production with a QCD matrix element (ME) calculation and leading-order (LO) accuracy in the parton shower evolution. The \NNPDF[2.3lo] parton distribution function (PDF)~\cite{Ball:2012cx} set was used, and \PYTHIA internal parameter values were set according to the A14 tune~\cite{ATL-PHYS-PUB-2014-021}. The \PYTHIA[8.230] MC sample is taken as the default choice to obtain the nominal result, as it has been tested extensively in previous ATLAS analyses~\cite{Aad_2016,ref20} and has been seen to accurately describe the data~\cite{ATL-PHYS-PUB-2019-017}.

Several alternative MC samples are used for the estimation of uncertainties coming from hadronisation modelling. Two sets of MC samples were produced using the \SHERPA[2.2.5]~\cite{Bothmann:2019yzt} generator, with the same ME for the \(2\rightarrow2\) process at LO, the same parton shower configurations based on Catani--Seymour dipole factorisation~\cite{Schumann:2007mg}, and the same \CT[14nnlo] PDF set~\cite{Dulat:2015mca}, but different hadronisation algorithms. The first set of samples uses the dedicated \SHERPA AHADIC model for hadronisation~\cite{Winter:2003tt}, based on cluster fragmentation. The second set of samples was generated with the same configuration but using the \SHERPA interface to the Lund string fragmentation model of \PYTHIA[6]~\cite{Sjostrand:2006za} and its decay tables.

Two multijet MC samples were generated at next-to-leading order (NLO) by \HERWIG[7.1.3]~\cite{Bellm:2015jjp} with the same \MMHT[nlo]~\cite{Harland-Lang:2014zoa} PDF set and hadronisation model, but with one using of the default parton shower model with angular ordering and the other using the dipole shower as an alternative, allowing an estimation of the effect of the shower model on the results.

Another set of QCD multijet samples with NLO precision in the ME is also included for the ME uncertainty estimation. They were generated with \POWHEGBOX[v2]~\cite{Nason:2004rx, Frixione:2007vw,Alioli:2010xd} and interfaced to \PYTHIA for the parton shower and hadronisation models. The \NNPDF[3.0nlo]~\cite{Ball:2014uwa} PDF set was used for these samples.

\FloatBarrier
\section{Object reconstruction and event selection}
\label{sec:Obj-event}

Jets are reconstructed with the \antikt algorithm~\cite{Cacciari:2008gp} with a radius parameter of $R=0.4$ using as constituents particle-flow (PFlow) objects~\cite{PERF-2015-09}, with the combined information from the ID and the calorimeter. 
Reconstructed jets are considered isolated if there is no other reconstructed jet within a cone of size $\Delta R = 0.7$ around the jet axis. Only isolated jets are considered in this study. An overall jet energy calibration~\cite{ATLAS:2020cli} is performed with a sequence of simulation-based corrections and \insitu calibrations, which accounts for residual detector effects as well as contributions from the effects of multiple simultaneous $pp$ collisions (\pileup). Only light-flavoured quark-jets ($u$, $d$, $s$) are considered in this study.

Tracks arising from charged particles are reconstructed~\cite{ATLAS:2017kyn} from the hits in the ID and are required to have transverse momentum $\pt > 500$~\MeV, $\abseta < 2.5$, at least one pixel hit and at least six hits in the silicon microstrip tracker, as well as transverse and longitudinal impact parameters with respect to the hard-scattering vertex that satisfy $|d_0| < 1$~mm and $|\zzsth| < 1$~mm respectively. Additionally, the event is required to have at least one vertex with two or more associated tracks. The vertex with the highest $\pt^2$ sum of the associated tracks is considered to be the primary vertex. The ghost-association technique~\cite{CACCIARI2008119} is employed to match tracks to jets; tracks are treated as four-vectors of infinitesimal magnitude during the jet reconstruction and are then assigned to the jet with which they are clustered.

A second jet collection used is called \enquote{truth jets}~\cite{ATLAS:2020cli}. Truth jets are reconstructed from stable final-state particles from the simulation samples, using the same \antikt $R=0.4$ algorithm as PFlow jets. They are geometrically matched to PFLow jets by requiring that their angular separation satisfies $\Delta R< 0.4$. Truth jets are assigned a flavour label~\cite{Aad_2014,ATL-PHYS-PUB-2017-009}, called \enquote{truth label}. The truth label of a jet is defined by the flavour of the highest-energy parton in the parton shower, before hadronisation, within a cone of size $\Delta R = 0.4$ around the jet axis. Using this definition, jets that originate from gluons splitting into $b$-quark or $c$-quark pairs are labelled as heavy-flavour jets, which are often identified by the presence of long-lived or leptonically decaying hadrons, and thus no special discriminant for heavy-flavour quarks is used here~\cite{CDF:2008ixu,ATLAS:2013uet}. Jets remain unlabelled if no truth parton with $\pt > 1$~\GeV\ is found within the cone surrounding the truth jet. Unlabelled jets typically arise from \pileup, and are less than 1\% of the dataset at $\pt > 500$~\GeV. They are thus ignored~\cite{ATLAS:2012mwf}.

Events used in the analysis were selected with a single-jet trigger, and must have at least two jets with $\pt > 500$~\GeV.
The two leading jets must each have $\abseta < 2.1$ to guarantee that they are well within the acceptance of the tracking detector.
The ratio of the leading jet's \pt to the sub-leading jet's \pt is required to be less than 1.5, to select a sample in which the two jets are balanced in \pt. The two leading-\pt jets are used to define quark-enriched and gluon-enriched subsamples.

For each selected jet pair, the jet with the higher \abseta value is selected to populate the quark-enriched sample. The other jet, which has a lower \abseta value, is assigned to a gluon-enriched sample. This selection strategy leverages the fact that in the high proton-momentum-fraction range, the PDF has a higher probability of including valence-quarks. Consequently, the ensemble of jets which are more forward (higher \abseta) have a higher probability of being quark-jets, whereas the ensemble of jets which are more central (lower \abseta) have a higher probability of being gluon-jets~\cite{ATLAS:2015rlw}. This is illustrated later for the \PYTHIA MC multijet sample, in Figure~\ref{fig:QG-Fmc}, where the quark-jet fraction is higher in the forward region than in the central region and also increases with jet \pt.

\FloatBarrier
\section{Tagger definitions}
\label{sec:QG-var}
Jet substructure variables are useful in developing $q/g$ taggers, given the predicted difference between the radiation patterns of quark- and gluon-jets. A variable well suited to this task is the track multiplicity, as gluon-jets are expected to have more constituents than quark-jets. Hence, the \ntrk in a jet can be used to define a single-variable $q/g$ tagger by imposing a requirement on its value. In this analysis, a more advanced $q/g$ tagger based on a BDT is also developed, using information about the jet \pt, \ntrk, the jet track width \wtrk~\cite{Aad_2014,PhysRevLett.110.212001}, and the two-point energy correlation function \cbeta~\cite{Moult:2016cvt,Larkoski:2013eya}, which takes into account the energy distribution within the jet.\footnote{The jet track width is defined as $\wtrk = \frac{ \sum_{\mathrm{track}\in\mathrm{jet}}\pttrk\Delta R_{\mathrm{track,jet}} }{ \sum_{\mathrm{track}\in\mathrm{jet}}\pttrk }$,
where \pttrk is the \pt of a charged track associated to the jet. The two-point energy correlation function instead is defined as $\cbeta = \frac{ \sum_{i, j\in\mathrm{jet}}^{i\neq j} p_{\mathrm{T}, i}p_{\mathrm{T}, j} \left( \Delta R_{i,j} \right)^{\beta=0.2} }{ \left( \sum_{\mathrm{track}\in\mathrm{jet}} \pttrk  \right)^{2} }$,
where $i$ and $j$ denote tracks associated with the jet and the sum runs over all the combinations of two tracks. 
The parameter $\beta$ is fixed to $0.2$, which is known to be suitable for $q/g$ tagging.} The \abseta of a jet is not used as an input to the BDT, as it could interfere with the definition of the quark- and gluon-enriched samples and distort the estimation of the fractions of quark- and gluon-jets using the method presented in Section~\ref{sec:QG-method}.

The BDT-tagger is trained using 60~million events with two jets from \PYTHIA MC samples described in Section~\ref{sec:Data-MC}, with a dataset distribution of 8\,:\,1\,:\,1 for training, validation, and testing, respectively. For the simulated events employed in the BDT training process, an additional processing step is implemented to obtain a flattened distribution of the \pt spectra for quark- and gluon-jets.
This step aims to emphasise the training for jets in the tail of the \pt spectrum, and to equalise the numbers of quark- and gluon-jets in the training. In training procedure, the LGBMClassifier from the lightGBM~\cite{NIPS2017_6449f44a} framework is used, with Optuna~\cite{akiba2019optuna} for hyperparameter tuning. A score is assigned to each BDT that goes into the boosting process based on its error rate. After 100 iterations of this procedure, a stable BDT is established, defined with 224 leaves. The BDT score is used to classify a jet as a quark-jet or a gluon-jet.
 
The performance of a jet tagger is evaluated using a receiver operating characteristic (ROC) curve defined from the quark- and gluon-jet efficiencies. The area under the ROC curve (AUC) is used as a metric to quantify the effectiveness of a tagger, with a larger AUC value indicating better performance. Figure~\ref{fig:QG-ROC_500_600} shows the performance of jet tagging variables. The BDT performs better than individual jet-substructure variables, meaning that the BDT-tagger can reject more gluon-jets than the \ntrk-only tagger at the same quark-jet efficiency. Since the tagging variables strongly depend on the \pt of a jet, performances and comparisons among taggers are given in different jet-\pt bins with boundaries at 500, 600, 800, 1000, 1200, 1500 and 2000~\GeV.

\begin{figure}[htbp]
\centering
\includegraphics[width=0.5\textwidth]{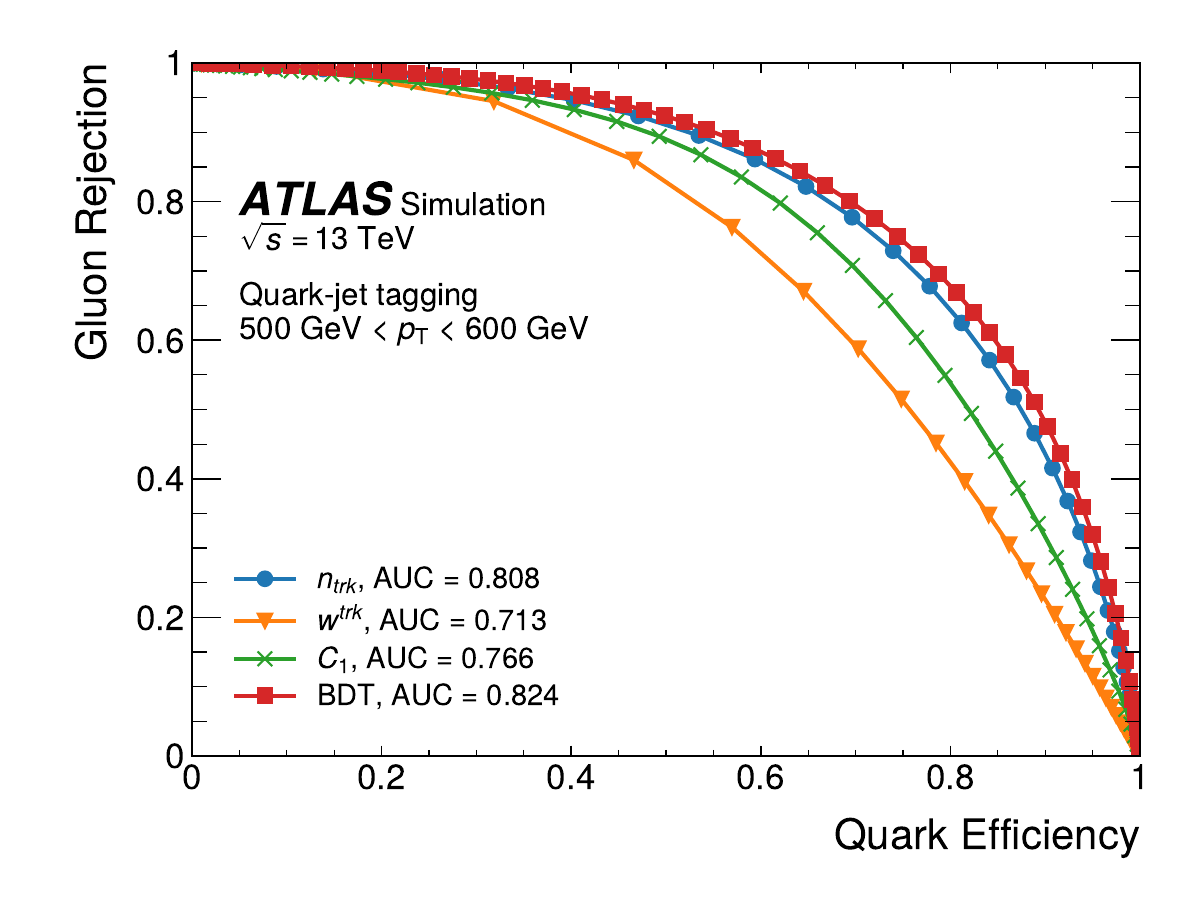}
\caption{ROC curves for quark-jet tagging variables, using the \PYTHIA MC sample.}
\label{fig:QG-ROC_500_600}
\end{figure}

\FloatBarrier

\section{Matrix method}
\label{sec:QG-method}
Evaluating the performance of the $q/g$ taggers under study needs samples containing solely quark- or gluon-jets. To extract the $q/g$ tagging-variable distribution shapes for quark- and gluon-jets in the data, a method that exploits samples with different $q/g$ fractions is used, called the matrix method~\cite{ATL-PHYS-PUB-2017-009}.

In the matrix method, the distribution of a jet variable $x$ for forward jets, $p_{\mathrm{F}}(x)$, and for central jets, $p_{\mathrm{C}}(x)$, can be written as:
\begin{eqnarray}
\label{eq:QG-matrix2}
\left(\begin{array}{c}
p_{\mathrm{F}}(x)\\
p_{\mathrm{C}}(x)\\
\end{array}\right)
&=&
\underbrace{
\left(\begin{array}{cc}
f_{\mathrm{F,Q}}&f_{\mathrm{F,G}}\\
f_{\mathrm{C,Q}}&f_{\mathrm{C,G}}\\
\end{array}\right)
}_{\scalebox{1}{$\equiv F$}}
\left(\begin{array}{c}
p_\mathrm{Q}(x)\\
p_\mathrm{G}(x)\\
\end{array}\right).
\end{eqnarray}

Here $p_{\mathrm{Q}}(x)$ and $p_{\mathrm{G}}(x)$ are the distributions of the variable $x$ for pure quark- and gluon-jets, respectively, and the matrix $F$ contains the fractions of quark- or gluon-jets in the samples of jets in the forward/central region. Such fractions are taken from MC simulation and are shown in Figure~\ref{fig:QG-Fmc} for the \PYTHIA MC samples. The matrix method allows $p_{\mathrm{Q}}(x)$ and $p_{\mathrm{G}}(x)$ to be extracted by inverting the matrix $F$ for every \pT range considered.

\begin{figure}[htb]
\centering
\subfloat{\label{fig:QG-Fmcd}\includegraphics[width=0.5\textwidth]{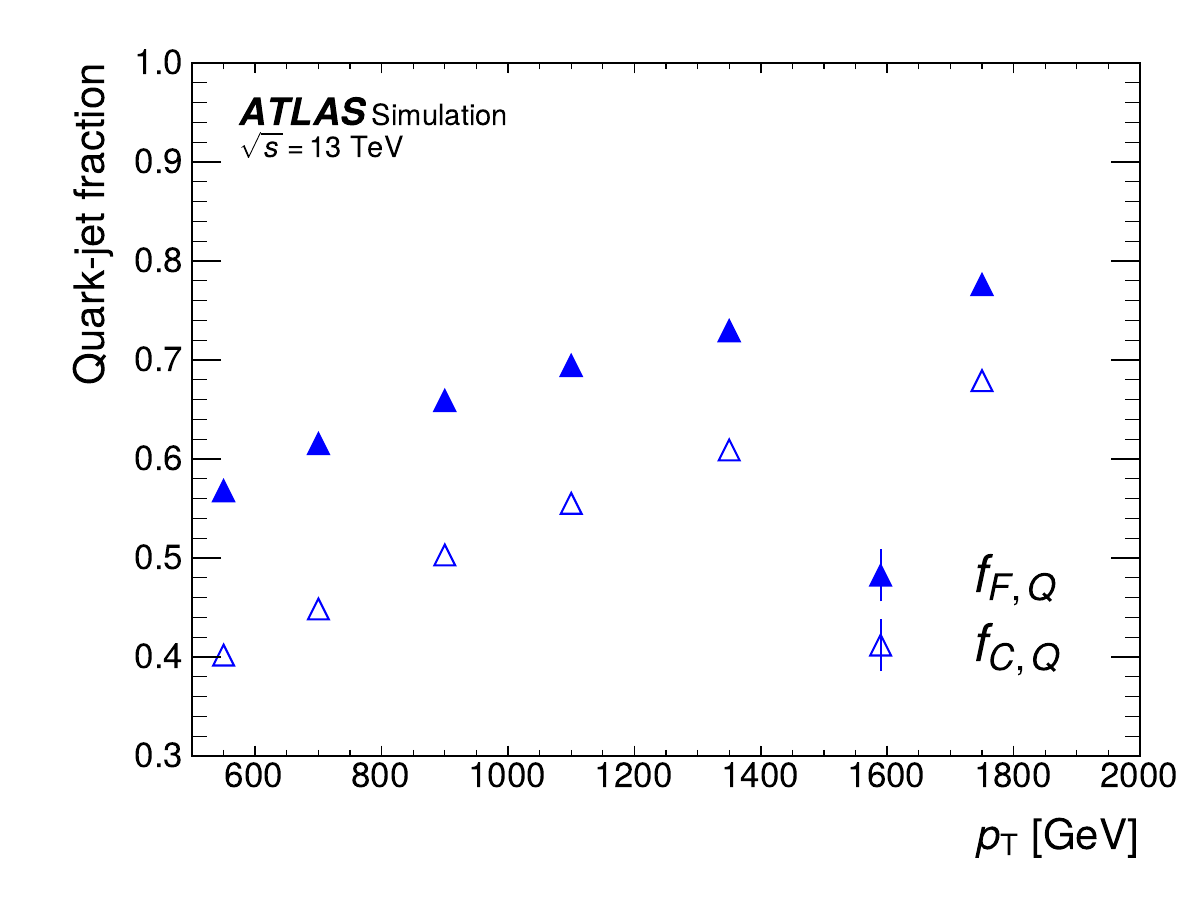}}\quad
\caption[]{
Fractions of quark-jets in the forward (triangles) and the central regions (open triangles) from the \PYTHIA MC multijet sample. The statistical uncertainty is smaller than the marker size.
\label{fig:QG-Fmc}
}
\end{figure}
 
\FloatBarrier

 
\label{sec:closure}
 
Equation~(\ref{eq:QG-matrix2}) is valid if it is assumed that the shapes of $p_{\mathrm{Q}}(x)$ and $p_{\mathrm{G}}(x)$ do not depend on whether the jets are in the central region or the forward region. Jet fragmentation at a $pp$ collider is expected to be mainly governed by the jet \pt and is generally considered independent of $\eta$ in accordance with the parton type. Therefore, an approach to extract distributions derived from the quark-jets' and gluon-jets' radiation patterns should be valid at the particle level. At the detector level, however, the measured radiation pattern inside jets is no longer independent of $\eta$, since changes in detector material and technology may cause variations in the response and introduce differences between the central and forward regions. These effects result in a non-closure of the matrix method.

A re-weighting procedure is applied to accommodate this feature and to ensure that the distributions of the jet tagging variables in the central and forward regions match. For each event, the central jet is weighted by a so-called re-weighting factor:

\begin{eqnarray*}
w(x)  =\frac{ p_\text{F} (x)}{ p_\text{C} (x)}.
\label{eq:QG-reweight}
\end{eqnarray*}
 
Even if the re-weighting factor corrects for an effect that is, at first order, independent on the origin of the jet, $w(x)$ can be calculated separately for truth-labelled quark-jets and gluon-jets. By default, the re-weighting factor obtained from truth-labelled quark-jets is applied to both the quark-jets and gluon-jets, while the re-weighting factor obtained from truth-labelled gluon-jets is used as an alternative to evaluate the systematic uncertainty associated with the re-weighting procedure.

After re-weighting, the extracted $p_{\mathrm{Q}}(x)$ and $p_{\mathrm{G}}(x)$ distributions exhibit good agreement with the truth distributions, as shown in Figure~\ref{fig:QG-pythia-closure-before}. The shapes of the $p_{\mathrm{Q}}(x)$ and $p_{\mathrm{G}}(x)$ distributions extracted from the \PYTHIA MC samples are similar to those in data, with differences within 25\%, hence validating the method. A residual non-closure of a few percent still remains, as shown in the middle panel of Figures~\ref{fig:QG-pythia-closure-before}(a)--\ref{fig:QG-pythia-closure-before}(d), and is taken as an MC non-closure systematic uncertainty, as described in Section~\ref{sec:QG-syst}.

\begin{figure}[htbp]
\centering
\subfloat[]{\label{fig:QG-pythia-UncPythiaQa_Ntrk111}\includegraphics[width=0.45\textwidth]{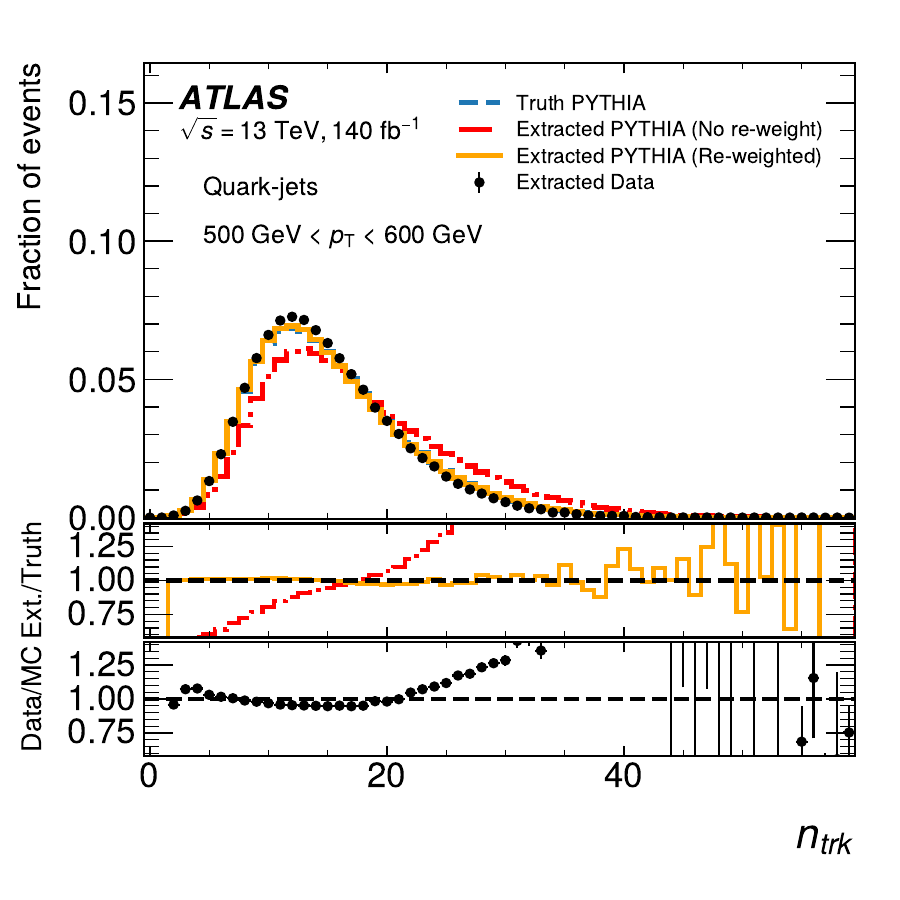}}\quad
\subfloat[
]{\label{fig:QG-pythia-UncPythiaQb_Ntrk111}\includegraphics[width=0.45\textwidth]{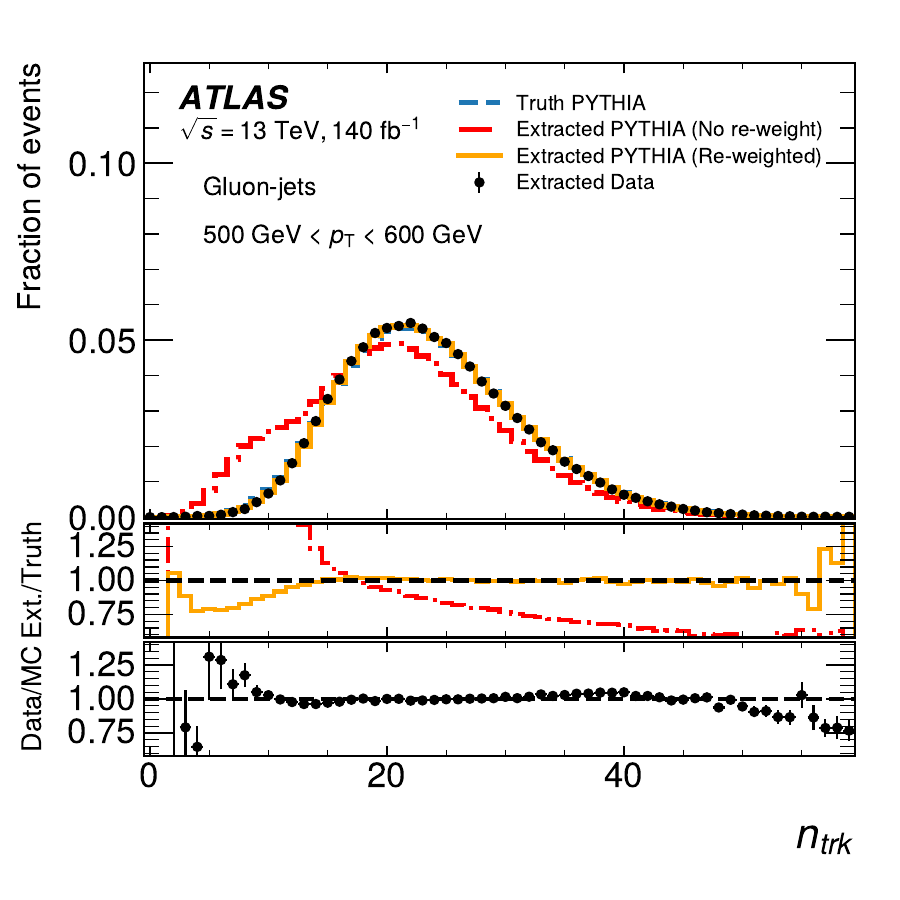}}\\
\subfloat[
]{\label{fig:QG-pythia-UncPythiaQa_Ntrk711}\includegraphics[width=0.45\textwidth]{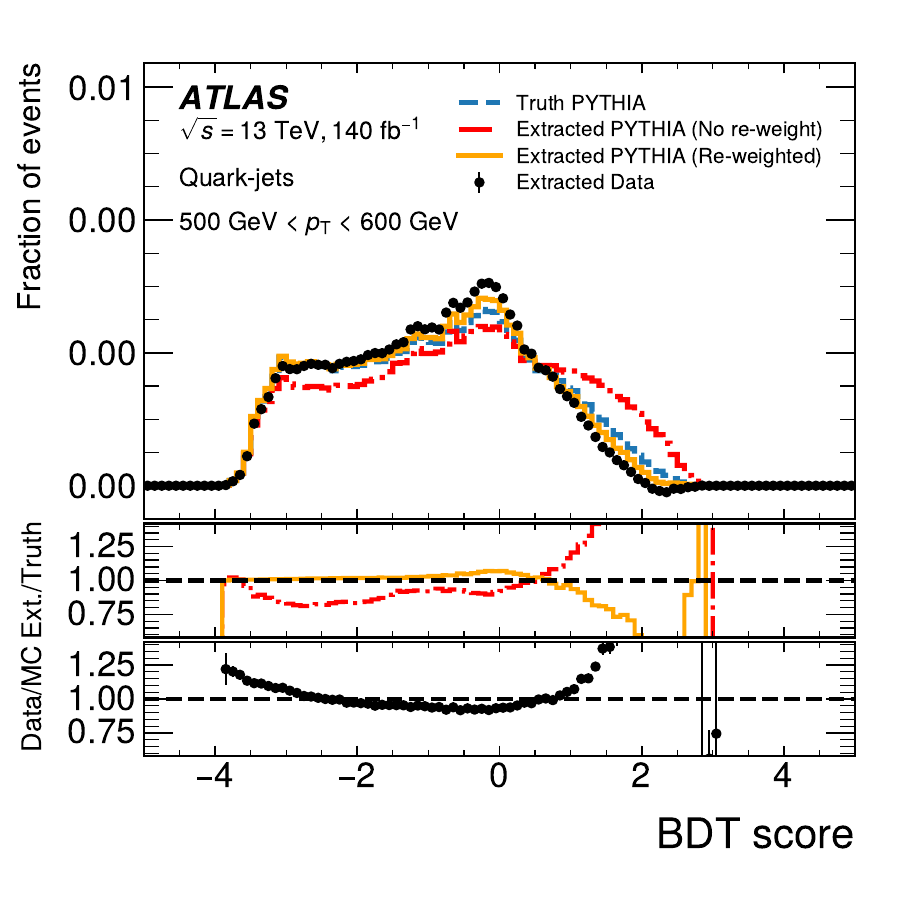}}\quad
\subfloat[
]{\label{fig:QG-pythia-UncPythiaQb_Ntrk811}\includegraphics[width=0.45\textwidth]{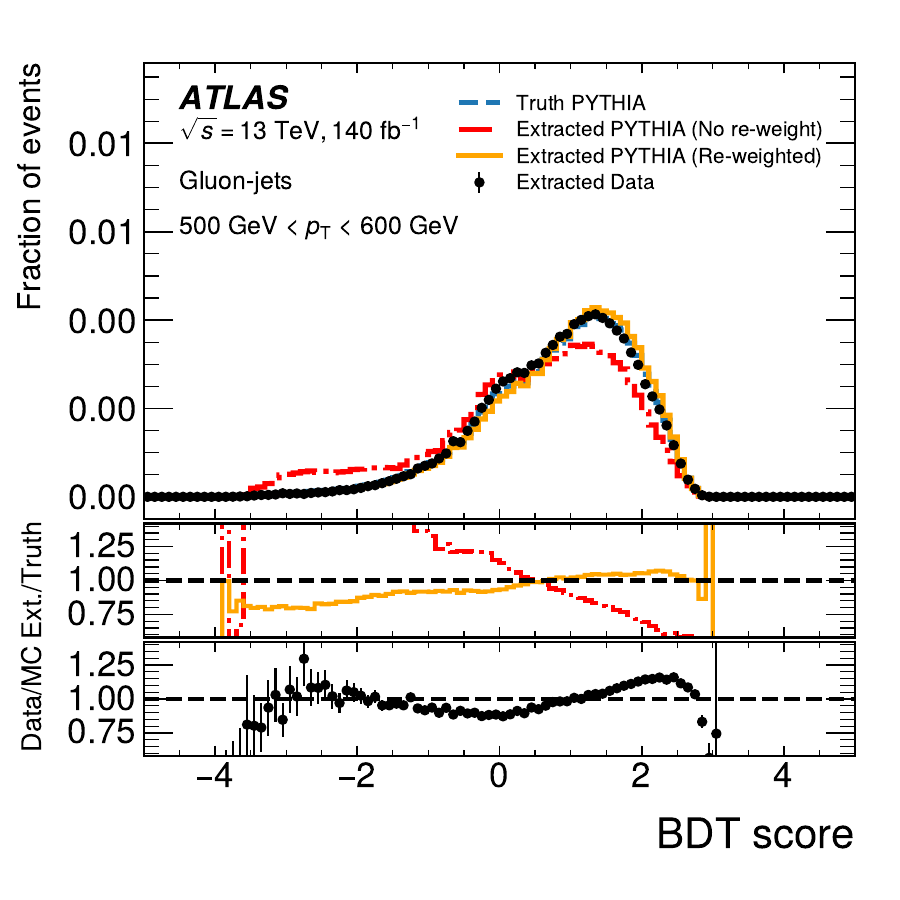}}
 
\caption[]{
The distributions of \ntrk (a,b) and BDT score (c,d) in quark-jets (left) and gluon-jets (right) in the \pt range between 500 \GeV\ and 600~\GeV\ obtained from truth \PYTHIA (dashed line), extracted \PYTHIA before re-weighting (dash dotted line) and after re-weighting (solid line), and extracted data (dots) are shown in the top panel. The middle panel shows the ratio of extracted \PYTHIA to truth \PYTHIA. The bottom panel shows the ratio of extracted data to extracted \PYTHIA after re-weighting.
\label{fig:QG-pythia-closure-before}
}
\end{figure}

The tagger working points (WP) are defined for fixed quark-jets efficiency in the nominal \PYTHIA MC sample, for both taggers. The efficiencies for quark- and gluon-jets at a given WP are defined as:
\begin{eqnarray}
\label{eq.q}
\epsilon_\text{Q/G} (x^{\text{WP}}) = \int_{x<x^\text{WP}} p_\text{Q/G} (x) \dif x.
\end{eqnarray}
 
Rejection factors for quark- and gluon-jets can also be defined, as:
\begin{eqnarray}
\label{eq.qq}
\xi_\text{Q/G} (x^{\text{WP}}) = 1 / \int_{x>x^\text{WP}} p_\text{Q/G} (x) \dif x  = 1 /  (1- \epsilon_\text{Q/G} (x^{\text{WP}})).
\end{eqnarray}
 
Differences between the quark-jet tagging efficiencies and gluon-jet rejection measured in data and the ones extracted from MC samples are described by data-to-MC scale factors (SF), for each $q/g$ tagger and in various \pt bins, at a fixed WP. The SF is defined using Eqs.~(\ref{eq.q}) and~(\ref{eq.qq}) for quark- and gluon-jets, respectively: 
\begin{eqnarray}
\textrm{SF}_{\mathrm{Q}} (x^{\text{WP}}) =\frac{\epsilon_\mathrm{Q}^\text{Data}(x^{\text{WP}})}{\epsilon_\mathrm{Q}^\text{MC}(x^{\text{WP}})}\\
\textrm{SF}_{\mathrm{G}} (x^{\text{WP}}) =\frac{\xi_\mathrm{G}^\text{Data}(x^{\text{WP}})}{\xi_\mathrm{G}^\text{MC}(x^{\text{WP}})},
\label{eq:QG-SF}
\end{eqnarray}
where $\epsilon_\mathrm{Q/G}^\text{Data} (x^{\text{WP}}) $ and $\epsilon_\mathrm{Q/G}^\text{MC} (x^{\text{WP}}) $ are $\epsilon_\mathrm{Q/G}(x^{\text{WP}})$ in data and MC samples, respectively. The same definitions apply to $\xi_\mathrm{Q/G}(x^{\text{WP}})$.
The WPs corresponding to 50\%, 60\%, 70\% and 80\% fixed quark-jets tagging efficiency have been studied and their corresponding SFs exhibit similar characteristics. The results for the 50\% WP are shown in Section~\ref{sec:QG-SF}.

\FloatBarrier

\FloatBarrier


\section{Systematic uncertainties}
\label{sec:QG-syst}

 
Several sources of systematic uncertainty affect the measurement of the SFs. Theoretical uncertainties arise from the modelling in the MC simulation, due to the choice of matrix element, parton showering model, PDF, renormalisation and factorisation scales, and hadronisation model. The experimental uncertainties coming from the calibration of the jet energy scale (JES) and jet energy resolution (JER)~\cite{PhysRevD.96.072002} and from track reconstruction are also taken into account. Uncertainties due to methodology, such as the one associated with the re-weighting procedure and the residual MC non-closure, are also considered and propagated to the final SF measurements.

\subsection{Theoretical uncertainty}
The uncertainty due to the modelling of the parton shower is obtained by comparing SFs in two \HERWIG~MC samples with the same ME and hadronisation model but different shower algorithms, as described in Section~\ref{sec:Data-MC}; this uncertainty varies between 1\% and 9\%. The systematic uncertainty due to the hadronisation modelling is estimated as the difference between the SFs obtained with two \SHERPA~MC samples with different hadronisation models; this uncertainty ranges between 1\% and 8\%. An additional uncertainty covering the calculation of the ME and its matching to the parton shower algorithm is estimated from the differences between the SFs extracted using \POWPY~and \PYTHIA~MC samples. This uncertainty amounts to approximately 1\% to 4\%.
 
The uncertainty due to the chosen PDF is evaluated using the LHAPDF recommendations~\cite{LHAPDF}. The uncertainty is estimated using the variations of the \NNPDF[2.3lo] PDF set in the nominal \PYTHIA~MC sample, and it amounts to 5\%--7\%.
 
Variations of the renormalisation (\muR) and factorisation (\muF) scales for initial- and final-state radiation are used as scale uncertainties to estimate the uncertainty due to missing higher-order corrections. Seven variations of (\muR,\,\muF), with their nominal values multiplied by factors of (0.5,\,0.5), (0.5,\,1), (1,\,0.5), (1,\,1), (2,\,1), (1\,,2) and (2,\,2), are used for the uncertainty, which is estimated from the envelope of the SFs obtained from such variations in \PYTHIA~MC samples. The uncertainty amounts to approximately 4\% to 7\%.
 
The splitting-Kernel variations~\cite{Mrenna:2016sih} pertain to modifications of the non-singular part of the splitting functions for initial-state radiation and final-state radiation, since significant uncertainties in the non-singular terms indicate more matched matrix elements included in the computation. The uncertainty is estimated by taking the envelope of the variations of non-singular terms and it is less than 1\%.
 
In total, the whole theoretical uncertainty amounts to approximately 18\% for both taggers and is found to be the main source of uncertainty.

\subsection{Experimental uncertainty}
Experimental uncertainties come from two sources: tracking efficiencies and the JES/JER calibration. The number of associated tracks is the most important input for both taggers, and the tracking-related systematic uncertainties can impact the SF measurements. The uncertainty in the number of reconstructed tracks is split into two terms: the uncertainty in the track reconstruction efficiency and the fake-track rate~\cite{ATLAS:2017kyn}. Both uncertainty sources are taken into account to recalculate the number of tracks associated with jets. The track reconstruction efficiency is affected by detector material uncertainties, which are the dominant source, and the physics model. They are estimated by comparing the track reconstruction efficiencies in MC samples with varied detector modelling. The fake-track rate uncertainty is estimated from a data-to-MC comparison of the evolution of the non-linear component of the track multiplicity as a function of the average mean number of interactions per bunch crossing. The tracking systematic uncertainty is obtained from changes in the SFs after applying the systematic variations and is approximately 1\% to 8\%.
 
The JES uncertainties~\cite{PhysRevD.96.072002} arise from calibrating the transverse momentum balance between central and forward jets, as well as accounting for single-particle and test-beam uncertainties. The JER uncertainties consider the JER difference between data and MC samples, by studying the dijet \pt balance asymmetry. An SF is obtained for each JES/JER variation, and the change from the nominal SF value is taken as the systematic uncertainty. The total JES/JER uncertainty is approximately 0.2\%.

\subsection{Methodological uncertainty}
Uncertainties associated with the matrix method come from the re-weighting process and the residual MC non-closure. In estimating the systematic uncertainty from the re-weighting process, the weights obtained from truth-labelled gluon-jets are used as an alternative, as explained in Section~\ref{sec:QG-method}. The resulting impact on the SFs is small (between 0.1\% and 0.5\%) across the whole \pt range considered. The residual MC non-closure, which is observed after the re-weighting procedure, affects the SFs at the 1\% level for both $q/g$ taggers studied.

The statistical uncertainty is calculated by varying the input data distributions bin-by-bin using a Poisson distribution with the number of events in each bin as the central value. The same procedure is applied to the MC samples, but using a Gaussian distribution. Each variation of the input distributions is used as an input to the matrix method. This procedure is repeated 5000 times, with the standard deviation of the uncertainties from all pseudo-datasets taken to be the statistical uncertainty of the scale factor. This uncertainty is approximately 0.1\%.

\FloatBarrier


\section{Results}
\label{sec:QG-SF}


Figure~\ref{fig:QG-pythia-Unc_Ntrk-wp} shows the gluon-jets efficiency factor defined by Eq.~(\ref{eq.q}) as a function of jet \pt in both the MC samples and data, for the 50\% quark-jets efficiency WP (50\% WP). For this WP, around 90\% of the gluon-jets are rejected by the \ntrk-only tagger, while approximately 93\% of gluon-jets are rejected by the BDT-tagger. The BDT-tagger is found to perform better or as well as the {\ntrk}-only tagger, i.e.\ it has a lower gluon-jet efficiency at the same WP. This is because the BDT-tagger includes more jet substructure variables. The difference between the gluon-jets efficiency in data and MC samples increases with increasing jet \pt, which is related to the MC modelling of gluons being different from the actual data. Such effect is more significant for the BDT-tagger.

\begin{figure}[htbp]
\centering
\includegraphics[width=0.45\textwidth]{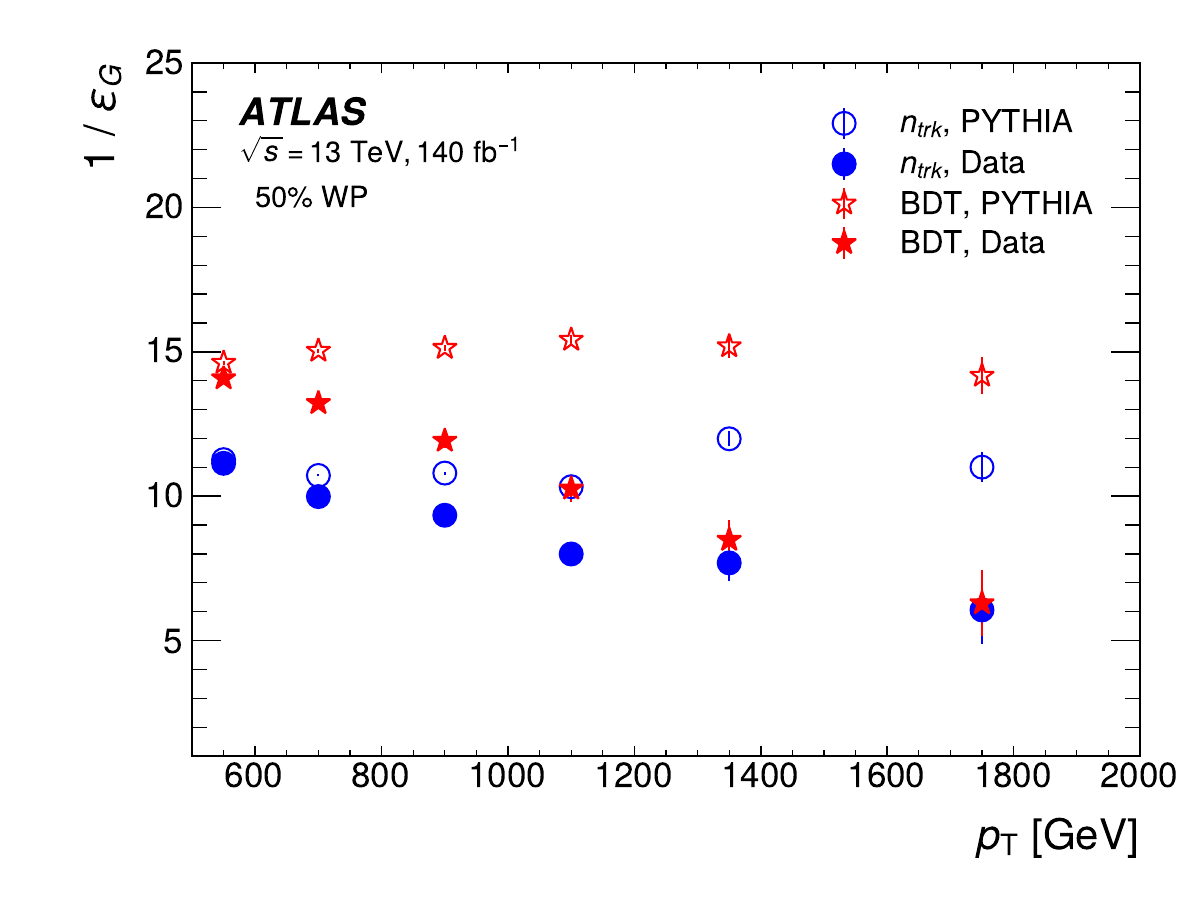}
\caption[]{
Inverse of the gluon-jet efficiency for the \ntrk-only tagger~(circles) and BDT-tagger~(stars) as a function of jet \pt at the 50\% WP in data (closed symbols) and the \PYTHIA MC sample (open symbols). The vertical error bars show the statistical uncertainty.
\label{fig:QG-pythia-Unc_Ntrk-wp}
}
\end{figure}

\begin{figure}[htbp]
\centering
\subfloat[]{\label{fig:QG-pythia-UncPythiaQa_Ntrk}\includegraphics[width=0.45\textwidth]{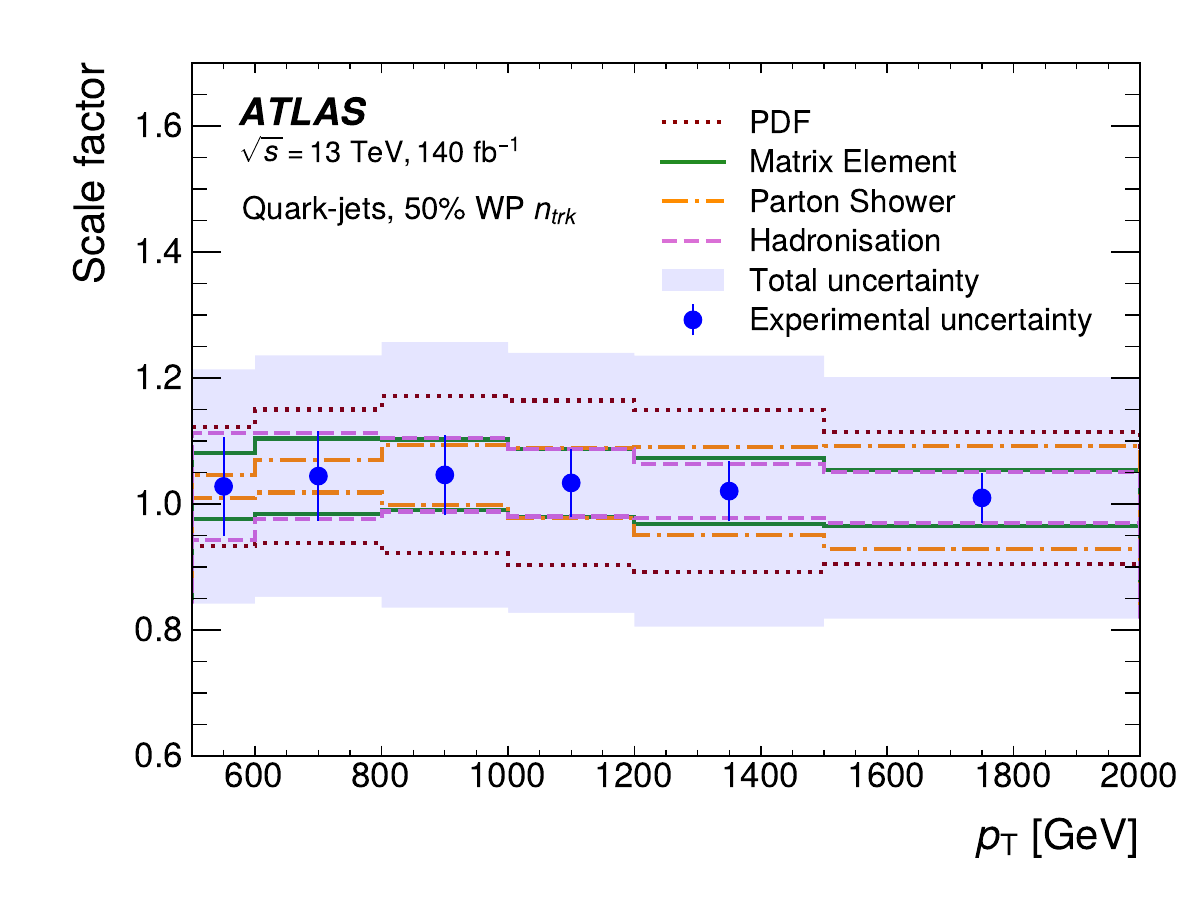}}	\quad
\subfloat[
]{\label{fig:QG-pythia-UncPythiaQb_Ntrk}\includegraphics[width=0.45\textwidth]{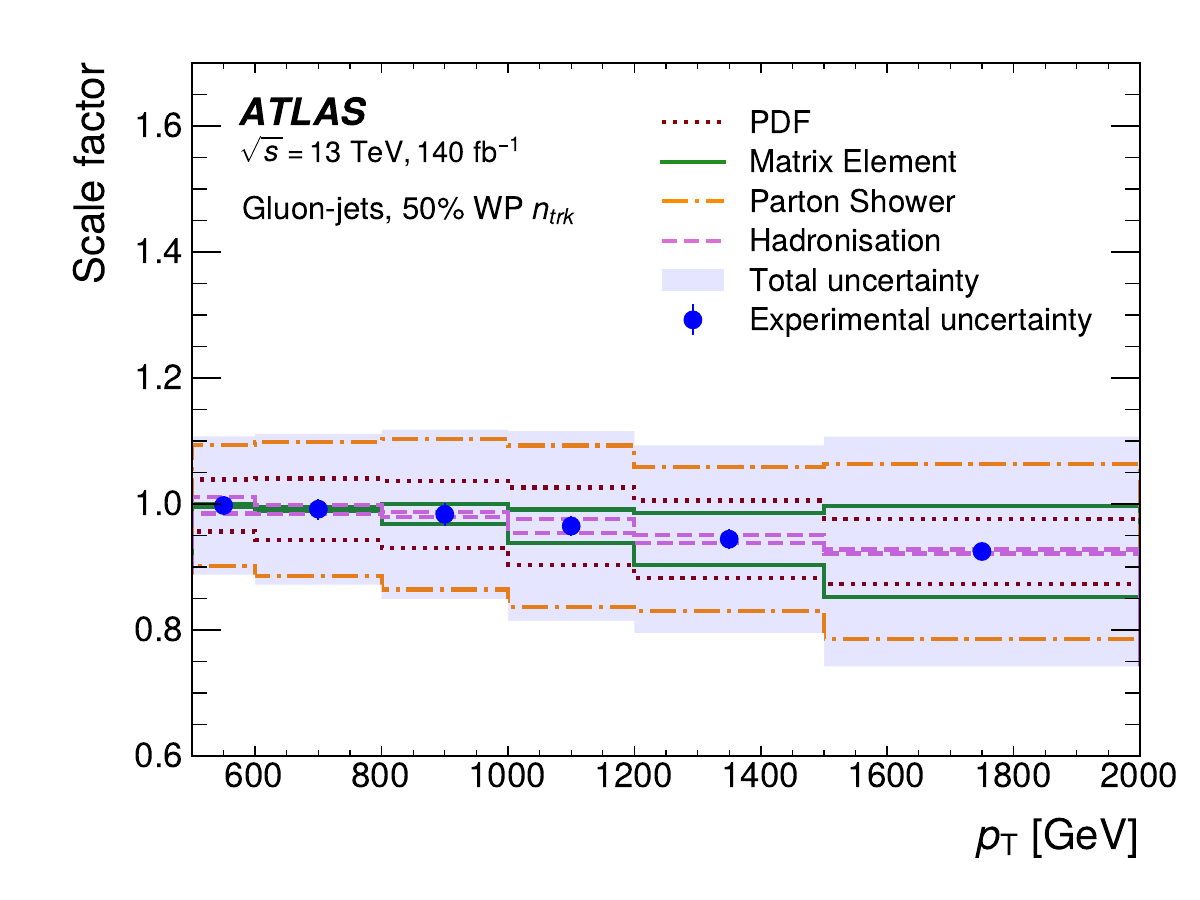}}\\
\subfloat[
]{\label{fig:QG-pythia-UncPythiaQa_Ntrk7}\includegraphics[width=0.45\textwidth]{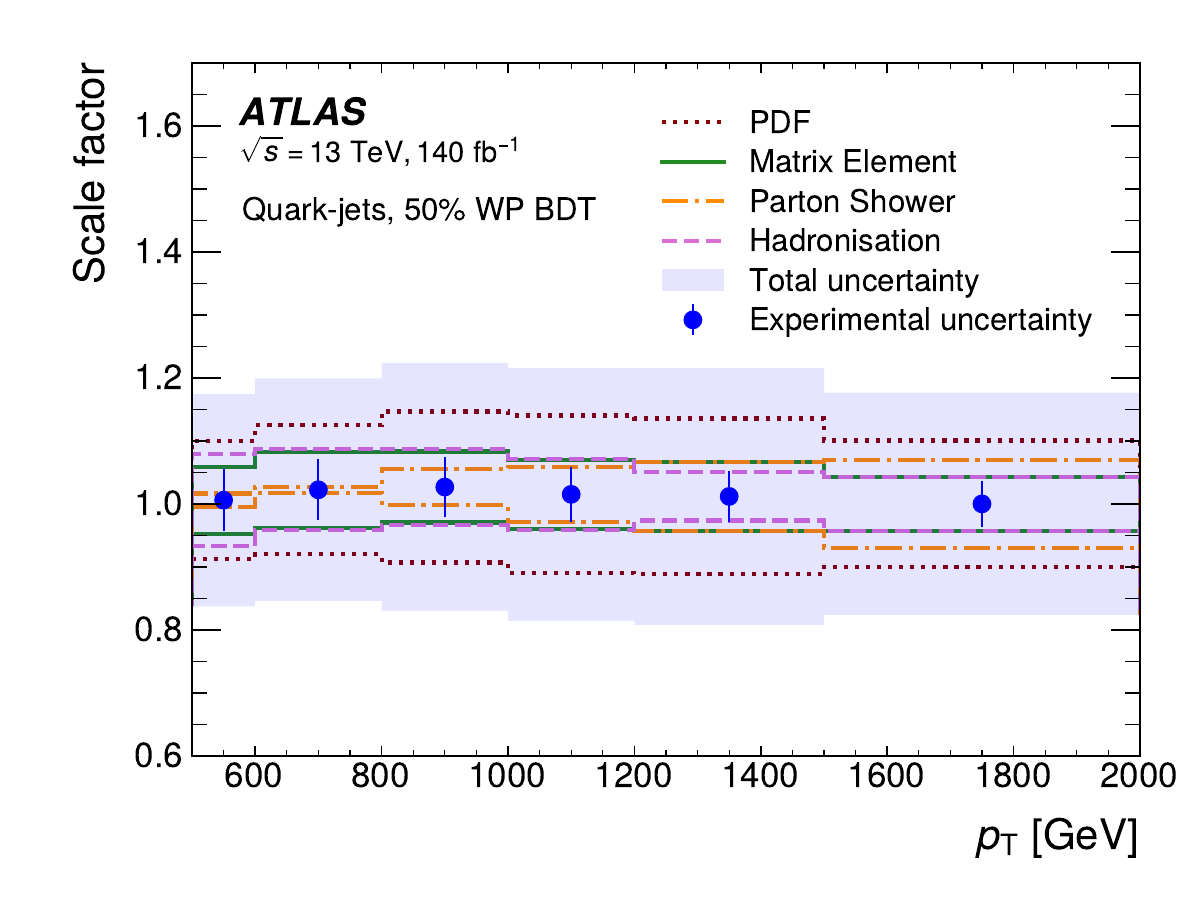}}\quad
\subfloat[
]{\label{fig:QG-pythia-UncPythiaQb_Ntrk8}\includegraphics[width=0.45\textwidth]{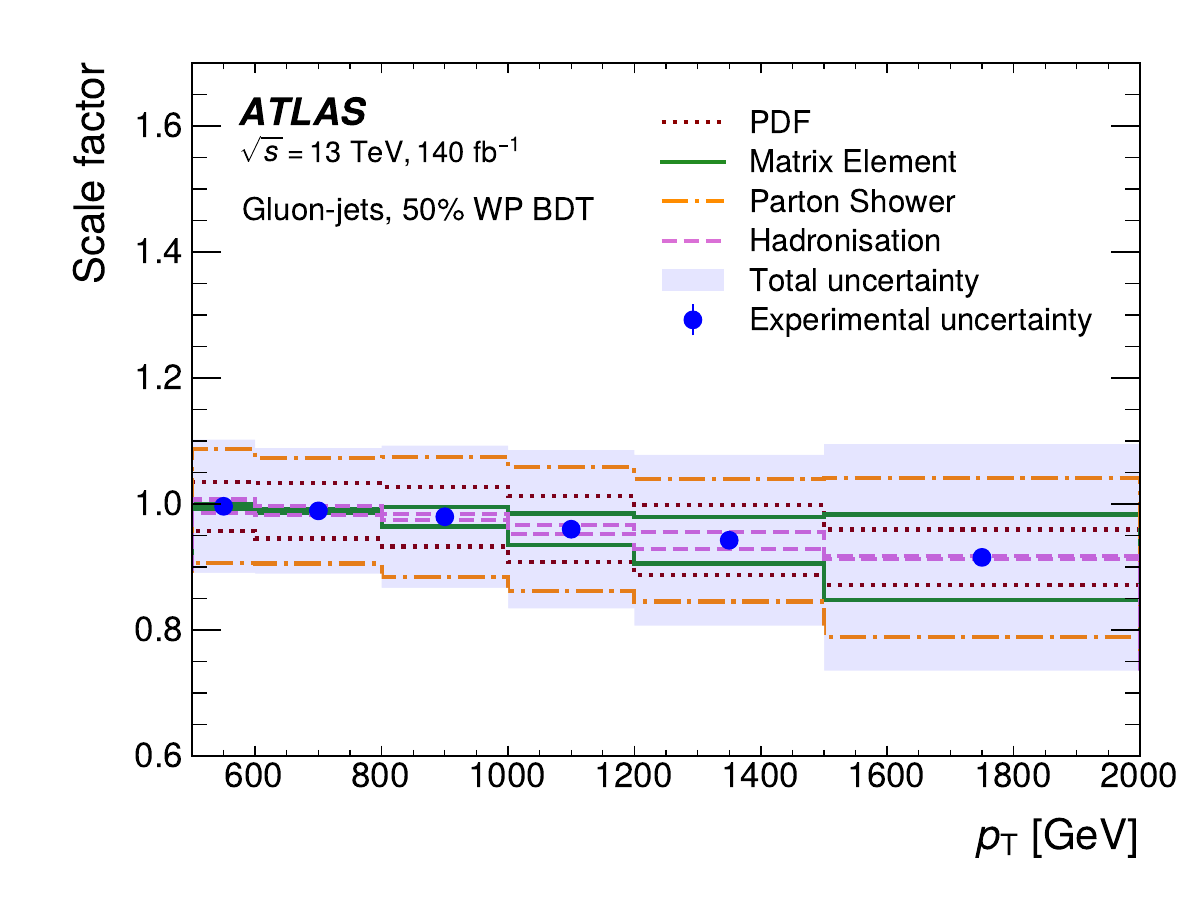}}
\caption[]{
The scale factors (dots) defined in Eq.~(\ref{eq:QG-SF}) with the total uncertainty (band), leading theoretical uncertainties (lines) and experimental uncertainty (vertical error bar) of the \ntrk-only tagger (a,b) and the BDT-tagger (c,d) as a function of jet \pt for quark-jets (left) and gluon-jets (right) at the 50\% WP using the \PYTHIA~MC sample.
\label{fig:QG-pythia-Unc_Ntrk}
}
\end{figure}

Figure~\ref{fig:QG-pythia-Unc_Ntrk} shows that the SFs for both quark-jets and gluon-jets at the 50\% quark-jets efficiency WP are between 0.92 and 1.02, with a total systematic uncertainty of about 20\%. The dominant source of systematic uncertainty is theoretical modelling. Tests were performed to check the stability of the results versus \abseta. The SF measurements were repeated after flattening the jet \abseta of the quark-/gluon-enriched subsamples. These alternative results are compatible with the nominal ones, within the total uncertainty reported.
 
Since analyses interested in using the results reported in Figure~\ref{fig:QG-pythia-Unc_Ntrk} may use different MC samples, a MC-to-MC SF is obtained by using each of the alternative MC samples and treating the \PYTHIA MC samples as pseudo-data, to account for modelling differences between the \PYTHIA and alternative MC samples. The MC-to-MC SFs for both jet taggers vary from 0.9 to 1.1 for most MC samples, as shown in Figure~\ref{fig:QG-pythia-Unc_gen}. There are relatively large gluon modelling differences between the \HERWIG dipole parton shower and the \PYTHIA parton shower, resulting in large  MC-to-MC SFs.

\begin{figure}[htbp]
\centering
\subfloat[]{\label{fig:QG-pythia-UncPythiaQa_Ntrk1}\includegraphics[width=0.45\textwidth]{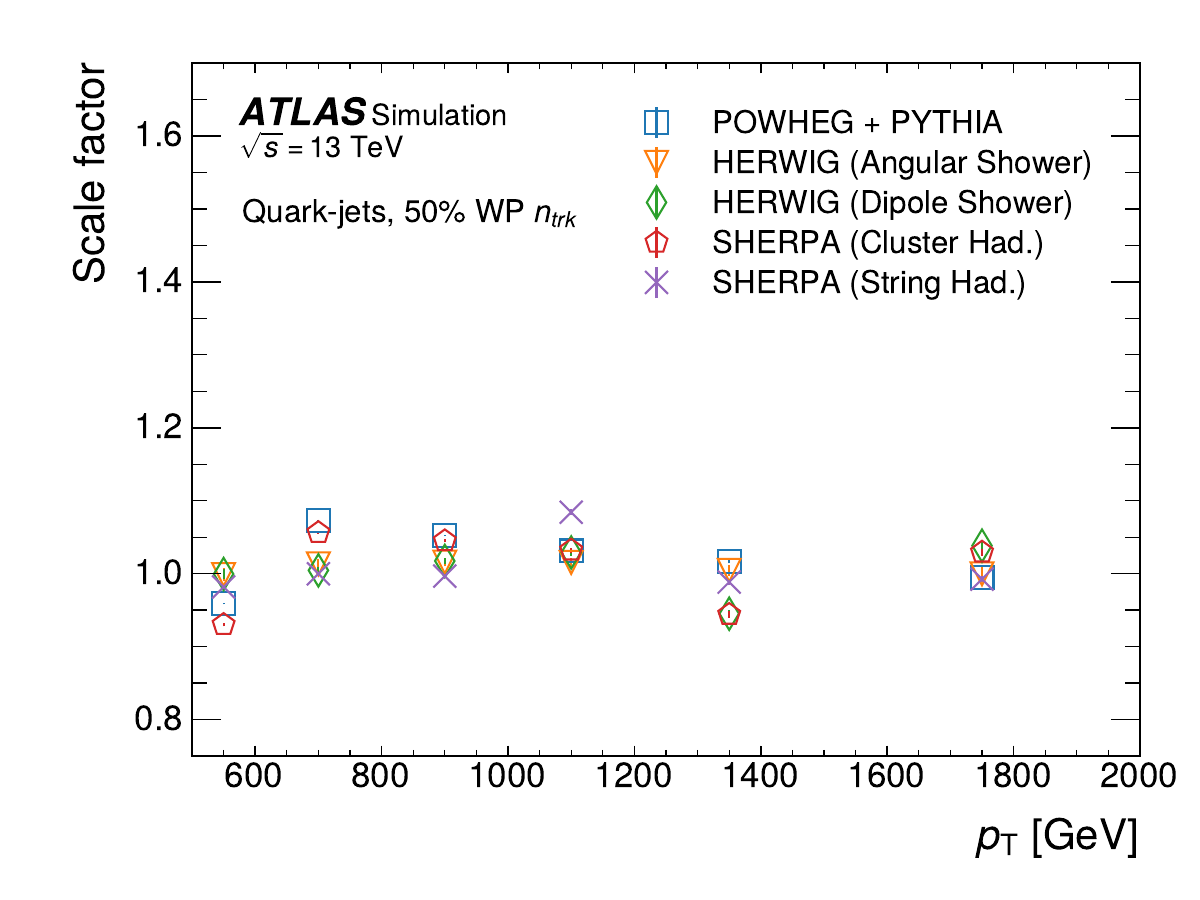}}\quad
\subfloat[
]{\label{fig:QG-pythia-UncPythiaQb_Ntrk1}\includegraphics[width=0.45\textwidth]{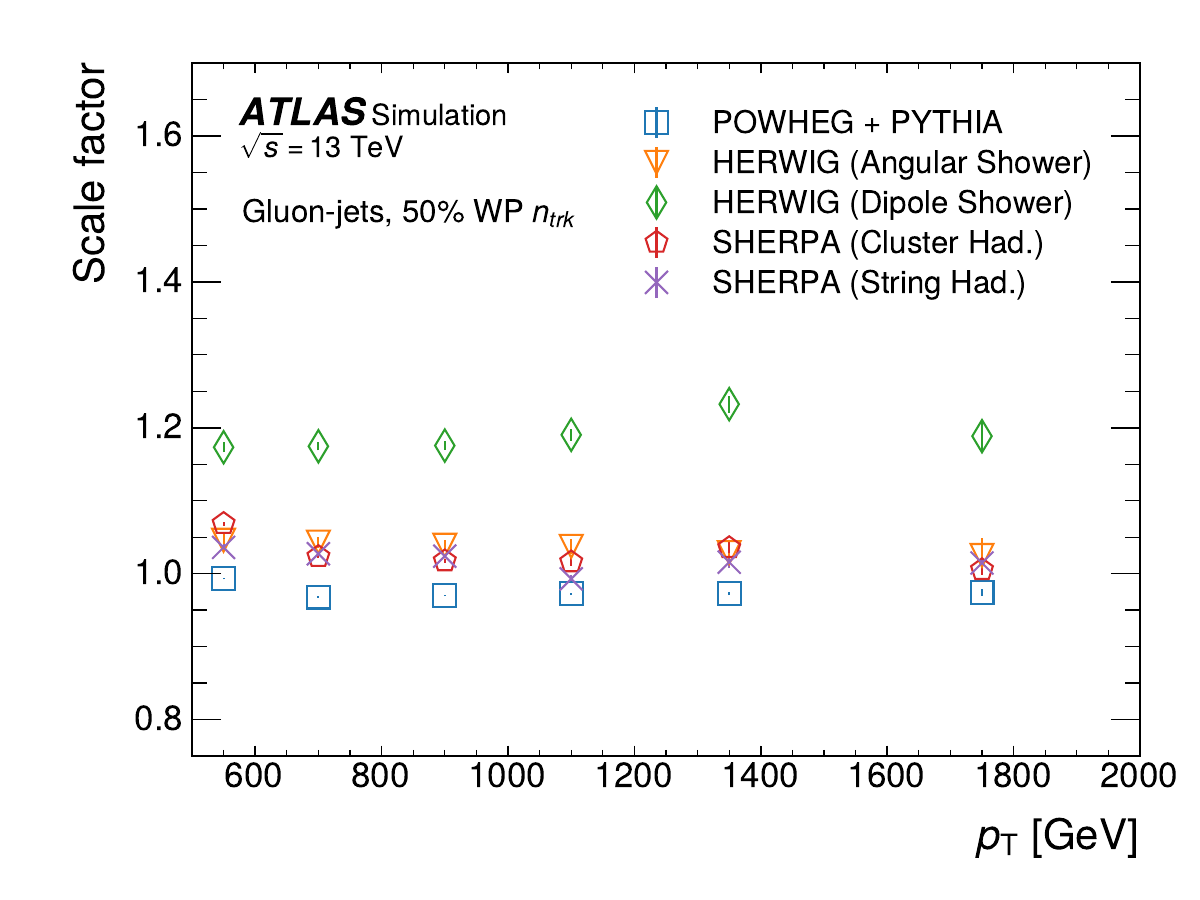}}\\
\subfloat[
]{\label{fig:QG-pythia-UncPythiaQa_Ntrk71}\includegraphics[width=0.45\textwidth]{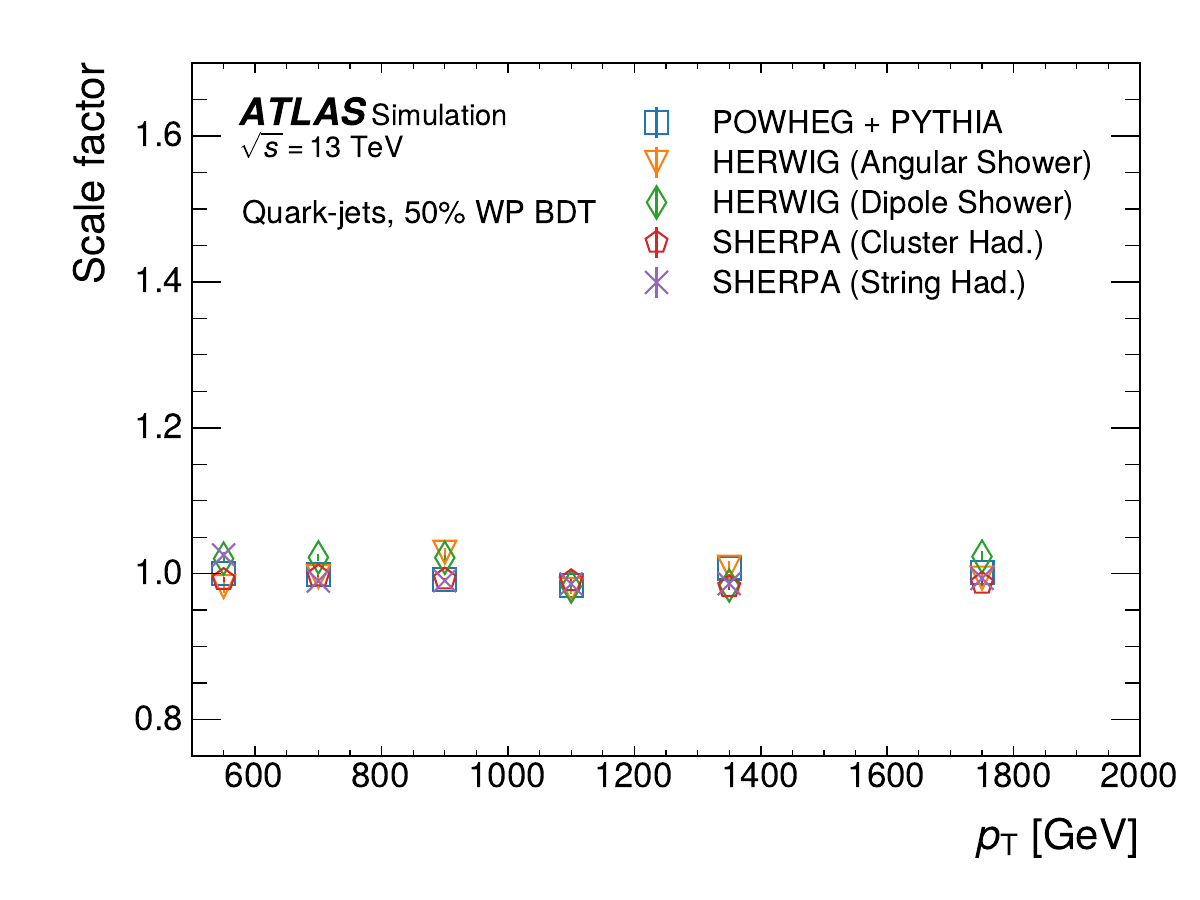}}\quad
\subfloat[
]{\label{fig:QG-pythia-UncPythiaQb_Ntrk81}\includegraphics[width=0.45\textwidth]{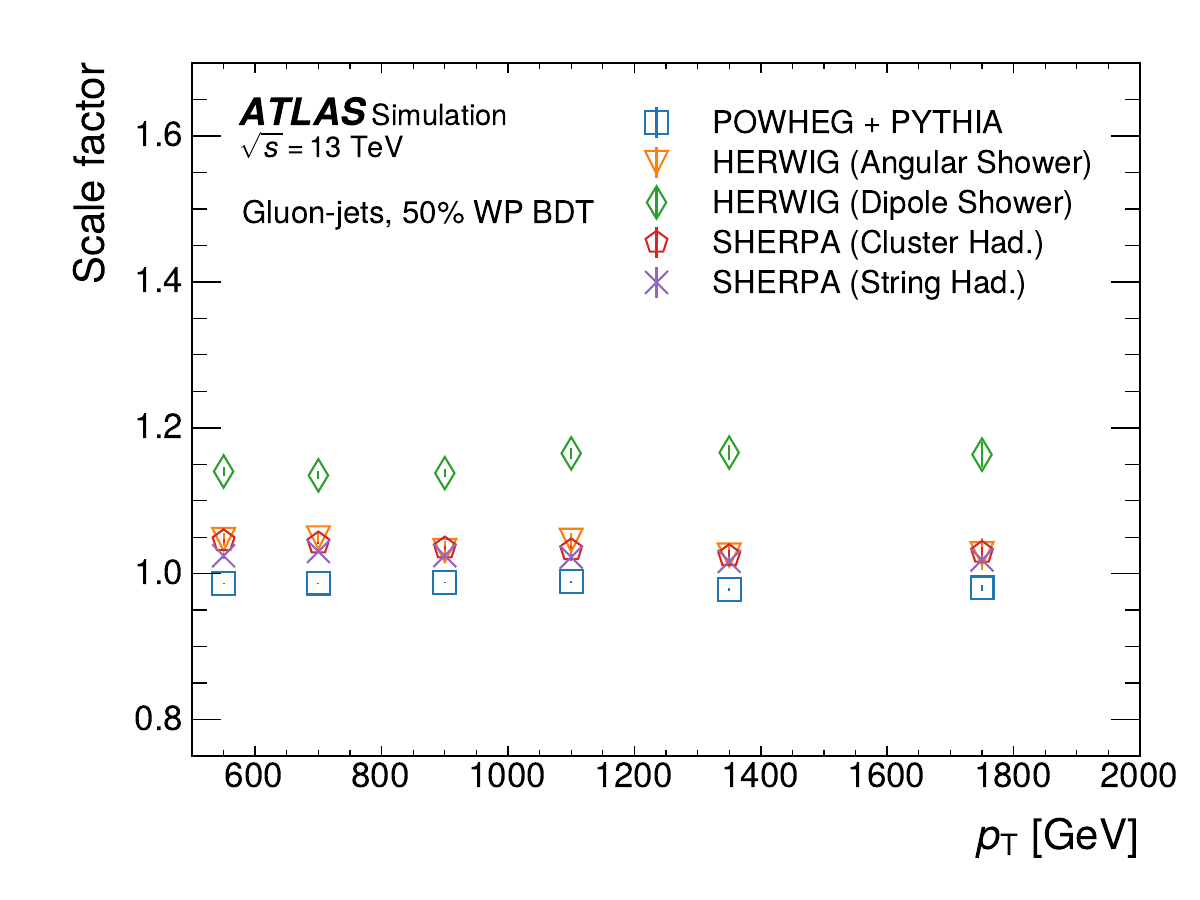}}
\caption[]{
The MC-to-MC scale factor for the \ntrk-only tagger (a,b) and BDT-tagger (c,d) as a function of jet \pt for quark-jets (left) and gluon-jets (right) at the 50\% WP. The vertical error bars show the statistical uncertainty. 
\label{fig:QG-pythia-Unc_gen}
}
\end{figure}

\FloatBarrier


\section{Conclusion}
\label{sec:conc}
 
The performance of taggers for quark- and gluon-initiated jets is studied using 140 fb$^{-1}$ of data from $pp$ collisions at $\sqrt{s} = 13$~\TeV\ collected by the ATLAS detector at the LHC, taking full advantage of the large dataset recorded from 2015 to 2018 to extend the taggers' reach to high jet energy. Two methods of jet tagging are investigated: a BDT-tagger, which combines several jet substructure observables, and a tagger based on the charged-particle jet-constituent multiplicity \ntrk. A matrix method is used to estimate the distribution shape of the tagging variables for quark- and gluon-jets, by combining information from quark-enriched and gluon-enriched samples obtained from a selection of two-jet events with jet \pt ranging from 500~\GeV\ to 2~\TeV. The variables considered are found to be described adequately by the MC, as differences with respect to ones measured in data are found to be smaller than 25\%, in all the different regions defined. When tested in data, the BDT-tagger is found to have better performance than the \ntrk-only tagger in selecting quark-jets over gluon-jets between 500 GeV and 1200 GeV, while above this range, the performance of the two taggers is comparable. For a fixed quark-jet efficiency of 50\%, the \ntrk-only tagger is able to reject approximately 90\% of gluon-jets, while the BDT-tagger is able to reject approximately 93\% of gluon-jets. A measurement of tagger performance differences in data and MC samples is provided through the definition of data-to-MC scale factors. The scale factors are measured in different jet-\pt intervals and are found to range from 0.92 to 1.02, with a total uncertainty of around 20\% which increases at higher \pt. The main source of uncertainty comes from the different modelling choices in MC simulation and amounts to approximately 18\% for both taggers. To account for differences among various MC generators, MC-to-MC scale factors are also provided, ranging from 0.9 to 1.1 for most MC samples.
The $q/g$ taggers developed in this article and the measurement of their SFs will benefit various analyses such as SM measurements that rely on the correct identification of jet origins, or new physics searches by enhancing their sensitivity to the presence of new particles.

\section*{Acknowledgements}


We thank CERN for the very successful operation of the LHC, as well as the
support staff from our institutions without whom ATLAS could not be
operated efficiently.

We acknowledge the support of 
ANPCyT, Argentina; 
YerPhI, Armenia; 
ARC, Australia; 
BMWFW and FWF, Austria; 
ANAS, Azerbaijan; 
CNPq and FAPESP, Brazil; 
NSERC, NRC and CFI, Canada; 
CERN; 
ANID, Chile; 
CAS, MOST and NSFC, China; 
Minciencias, Colombia; 
MEYS CR, Czech Republic; 
DNRF and DNSRC, Denmark; 
IN2P3-CNRS and CEA-DRF/IRFU, France; 
SRNSFG, Georgia; 
BMBF, HGF and MPG, Germany; 
GSRI, Greece; 
RGC and Hong Kong SAR, China; 
ISF and Benoziyo Center, Israel; 
INFN, Italy; 
MEXT and JSPS, Japan; 
CNRST, Morocco; 
NWO, Netherlands; 
RCN, Norway; 
MEiN, Poland; 
FCT, Portugal; 
MNE/IFA, Romania; 
MESTD, Serbia; 
MSSR, Slovakia; 
ARRS and MIZ\v{S}, Slovenia; 
DSI/NRF, South Africa; 
MICINN, Spain; 
SRC and Wallenberg Foundation, Sweden; 
SERI, SNSF and Cantons of Bern and Geneva, Switzerland; 
MOST, Taipei; 
TENMAK, T\"urkiye; 
STFC, United Kingdom; 
DOE and NSF, United States of America. 
In addition, individual groups and members have received support from 
BCKDF, CANARIE, CRC and DRAC, Canada; 
PRIMUS 21/SCI/017 and UNCE SCI/013, Czech Republic; 
COST, ERC, ERDF, Horizon 2020, ICSC-NextGenerationEU and Marie Sk{\l}odowska-Curie Actions, European Union; 
Investissements d'Avenir Labex, Investissements d'Avenir Idex and ANR, France; 
DFG and AvH Foundation, Germany; 
Herakleitos, Thales and Aristeia programmes co-financed by EU-ESF and the Greek NSRF, Greece; 
BSF-NSF and MINERVA, Israel; 
Norwegian Financial Mechanism 2014-2021, Norway; 
NCN and NAWA, Poland; 
La Caixa Banking Foundation, CERCA Programme Generalitat de Catalunya and PROMETEO and GenT Programmes Generalitat Valenciana, Spain; 
G\"{o}ran Gustafssons Stiftelse, Sweden; 
The Royal Society and Leverhulme Trust, United Kingdom.

The crucial computing support from all WLCG partners is acknowledged gratefully, in particular from CERN, the ATLAS Tier-1 facilities at TRIUMF/SFU (Canada), NDGF (Denmark, Norway, Sweden), CC-IN2P3 (France), KIT/GridKA (Germany), INFN-CNAF (Italy), NL-T1 (Netherlands), PIC (Spain), RAL (UK) and BNL (USA), the Tier-2 facilities worldwide and large non-WLCG resource providers. Major contributors of computing resources are listed in Ref.~\cite{ATL-SOFT-PUB-2021-003}.


\printbibliography

\clearpage
 
\begin{flushleft}
\hypersetup{urlcolor=black}
{\Large The ATLAS Collaboration}

\bigskip

\AtlasOrcid[0000-0002-6665-4934]{G.~Aad}$^\textrm{\scriptsize 102}$,
\AtlasOrcid[0000-0002-5888-2734]{B.~Abbott}$^\textrm{\scriptsize 120}$,
\AtlasOrcid[0000-0002-1002-1652]{K.~Abeling}$^\textrm{\scriptsize 55}$,
\AtlasOrcid[0000-0001-5763-2760]{N.J.~Abicht}$^\textrm{\scriptsize 49}$,
\AtlasOrcid[0000-0002-8496-9294]{S.H.~Abidi}$^\textrm{\scriptsize 29}$,
\AtlasOrcid[0000-0002-9987-2292]{A.~Aboulhorma}$^\textrm{\scriptsize 35e}$,
\AtlasOrcid[0000-0001-5329-6640]{H.~Abramowicz}$^\textrm{\scriptsize 151}$,
\AtlasOrcid[0000-0002-1599-2896]{H.~Abreu}$^\textrm{\scriptsize 150}$,
\AtlasOrcid[0000-0003-0403-3697]{Y.~Abulaiti}$^\textrm{\scriptsize 117}$,
\AtlasOrcid[0000-0002-8588-9157]{B.S.~Acharya}$^\textrm{\scriptsize 69a,69b,m}$
% ...
\bigskip
\\

% ...
$^{29}$Physics Department, Brookhaven National Laboratory, Upton NY; United States of America.\\
% ...
$^{35}$$^{(a)}$Facult\'e des Sciences Ain Chock, R\'eseau Universitaire de Physique des Hautes Energies - Universit\'e Hassan II, Casablanca;$^{(b)}$Facult\'{e} des Sciences, Universit\'{e} Ibn-Tofail, K\'{e}nitra;$^{(c)}$Facult\'e des Sciences Semlalia, Universit\'e Cadi Ayyad, LPHEA-Marrakech;$^{(d)}$LPMR, Facult\'e des Sciences, Universit\'e Mohamed Premier, Oujda;$^{(e)}$Facult\'e des sciences, Universit\'e Mohammed V, Rabat;$^{(f)}$Institute of Applied Physics, Mohammed VI Polytechnic University, Ben Guerir; Morocco.\\
% ...
$^{49}$Fakult\"{a}t Physik , Technische Universit{\"a}t Dortmund, Dortmund; Germany.\\
% ...
$^{55}$II. Physikalisches Institut, Georg-August-Universit\"{a}t G\"ottingen, G\"ottingen; Germany.\\
% ...
$^{69}$$^{(a)}$INFN Gruppo Collegato di Udine, Sezione di Trieste, Udine;$^{(b)}$ICTP, Trieste;$^{(c)}$Dipartimento Politecnico di Ingegneria e Architettura, Universit\`a di Udine, Udine; Italy.\\
% ...
$^{102}$CPPM, Aix-Marseille Universit\'e, CNRS/IN2P3, Marseille; France.\\
% ...
$^{117}$Department of Physics, New York University, New York NY; United States of America.\\
% ...
$^{120}$Homer L. Dodge Department of Physics and Astronomy, University of Oklahoma, Norman OK; United States of America.\\
% ...
$^{150}$Department of Physics, Technion, Israel Institute of Technology, Haifa; Israel.\\
$^{151}$Raymond and Beverly Sackler School of Physics and Astronomy, Tel Aviv University, Tel Aviv; Israel.\\
% ...
$^{a}$ Also Affiliated with an institute covered by a cooperation agreement with CERN.\\
$^{b}$ Also at An-Najah National University, Nablus; Palestine.\\
$^{c}$ Also at Borough of Manhattan Community College, City University of New York, New York NY; United States of America.\\
$^{d}$ Also at Center for High Energy Physics, Peking University; China.\\
$^{e}$ Also at Center for Interdisciplinary Research and Innovation (CIRI-AUTH), Thessaloniki; Greece.\\
$^{f}$ Also at Centro Studi e Ricerche Enrico Fermi; Italy.\\
% ...
$^{m}$ Also at Department of Physics, King's College London, London; United Kingdom.\\
% ...

\end{flushleft}


\end{document}